\documentclass[11pt]{article}

\usepackage{graphicx,psfrag,epsf}
\usepackage{enumerate}
\usepackage{cite}
\usepackage[all,poly,ps,color]{xy}
\usepackage[utf8]{inputenc}
\usepackage[T1]{fontenc}
\usepackage{url,epsfig,amssymb}
\usepackage{color}
\usepackage{geometry}
\usepackage{sectsty}
\usepackage[normalem]{ulem}
\usepackage{amsmath}
\sectionfont{\sffamily\bfseries\upshape\large}
\subsectionfont{\sffamily\bfseries\upshape\normalsize}
\subsubsectionfont{\sffamily\bfseries\upshape\normalsize}
\makeatletter

\usepackage{hyperref}

\title{Fitting Bayesian item response models in Stata and Stan\footnote{For the {\em Stata Journal}.  We thank the Institute of Education Sciences for partial support of this work.}}
\author{Robert L. Grant\footnote{Kingston University \& St George's, University of London},
             Daniel C. Furr\footnote{University of California at Berkeley},
	      Bob Carpenter\footnote{Columbia University}, and
	      Andrew Gelman\footnote{Columbia University}}
\date{5 Dec 2016}

\begin{document}

\newcommand{\UpperRhatRaschStata} {$1.01$ } 
\newcommand{\UpperRhatHraschStata} {$1.01$ } 
\newcommand{\UpperRhatRaschStan} {$1.01$ } 
\newcommand{\UpperRhatHraschStan} {$1.01$ } 
\newcommand{\MinutesRaschStata} {$15.9$ } 
\newcommand{\MinutesHraschStata} {$16.6$ } 
\newcommand{\MinutesRaschStan} {$16.6$ } 
\newcommand{\MinutesHraschStan} {$24.1$ }

\maketitle

\begin{abstract}
Stata users have access to two easy-to-use implementations of Bayesian inference:  Stata's native {\tt bayesmh} function and StataStan, which calls the general Bayesian engine Stan.  We compare these on two models that are important for education research:  the Rasch model and the hierarchical Rasch model.  StataStan fits a more general range of models than can be fit by {\tt bayesmh} and uses a superior sampling algorithm:  Hamiltonian Monte Carlo using the no-U-turn sampler.  Further, StataStan can run in parallel on multiple CPU cores, regardless of the flavor of Stata. Given its advantages, and that Stan is open-source and can be run directly from Stata do-files, we recommend that Stata users who are interested in Bayesian methods consider using StataStan.
\end{abstract}

\section{Introduction}

Stata is a statistical software package that is popular in social science, economics, and biostatistics.  In 2015, it became possible to routinely fit Bayesian models in Stata in two different ways:  (a) the introduction of Bayesian modeling commands in Stata software version 14 \cite{stata14}, which use the Metropolis-Hastings algorithm and Gibbs sampler; and (b) the release of StataStan, an interface to the open-source Bayesian software Stan \cite{StataStan, stanmanual}. Previously, Bayesian methods were only available in Stata by user-written commands to interface with external software such as BUGS, JAGS, or MLwiN.

At the time of writing, the native Bayes implementation in Stata, {\tt bayesmh}, allows a choice among 10 likelihood functions and 18 prior distributions. The command is explicitly focused around regression models, and extensions to hierarchical (multilevel) models are possible with the inclusion of hyperpriors.  In addition, the user may write customized likelihood functions or customized posterior distributions.

Stan is an open-source, collaboratively-built software project for Bayesian inference which allows general continuous-parameter models, including all the models that can be fit in Stata's {\tt bayesmh} and many others. Stan has been applied to a wide range of complex statistical models including time series, imputation, mixture models, meta-analysis, cluster analysis, Gaussian processes and item-response theory. These extend beyond the current (Stata 14.1) capability of \texttt{bayesmh}.  Stan can run from various data analysis environments such as Stata, R, Python, and Julia, and also has a command-line interface (CmdStan). Stan uses Hamiltonian Monte Carlo (HMC) and the no-U-turn sampler (NUTS) \cite{hoffman} with additional options of variational inference \cite{kucukelbir} and the L-BFGS optimization algorithm \cite{nocedal}. The advantages of HMC and NUTS in speed, stability with regard to starting values, and efficiency over Metropolis-Hastings and Gibbs have been described elsewhere \cite{neal, hoffman}. As a result of the Hamiltonian dynamics, HMC is rotation-invariant, making it well-suited to highly correlated parameters. It is also not slowed down by non-conjugate models.

The languages used by these packages are notably different. In Stan, models are specified in a series of probability statements specifying prior distributions and likelihoods. {\tt bayesmh} follows standard Stata syntax to give a compact specification of the most common regression and related models. Stan works by translating the user's model code into C++, then compiling and running the resulting executable file. Stan can run in parallel on multi-core computers, provided that the number of available cores was specified when installing CmdStan itself.

In the present paper, we compare {\tt bayesmh} and StataStan on some item response models.  These logistic regression, or Rasch, models, are popular in education research and in political science (where they are called ideal-point models) \cite{rasch1960}.

\section{Models}

We fitted the models using data simulated as specified below.  We checked that the {\tt bayesmh} and StataStan implementations gave the same answer (modulo the inevitable Monte Carlo error of these stochastic algorithms) and we then compared the programs on speed and efficiency in terms of time per the number of effective independent samples.

The Rasch model can be written as,
\begin{equation}
	\mathrm{Pr} (y_{ip} = 1 | \theta_p, \delta_i) =
	\mathrm{logit}^{-1} (\theta_p - \delta_i)
\end{equation}
\begin{equation}
	\theta_p \sim \mathrm{N} (0, \sigma^2),
\end{equation}
where $y_{ip}=1$ if person $p$ responded to item $i$ correctly and is 0 otherwise, and $i,p \in \mathbb{N}$; $1 \leq i \leq I$; $1 \leq p \leq P$.  The parameter $\theta_p$ represents the latent ``ability'' of person $p$, and $\delta_i$ is a parameter for item $i$. We considered a simple version of the model in which the abilities are modeled as exchangeable draws from a normal distribution with scale $\sigma$. We assigned a $\mbox N(0,10)$ prior distribution to $\delta_i$ and took two approaches to priors for $\sigma$. Firstly, we matched the Rasch model example in the Stata 14 manual \cite{stata14}, which uses an inverse-gamma prior for $\sigma^2$, which we do not recommend \cite{priors}. Secondly, we used a preferred approach of uniform priors for $\sigma$, which is the Stan default if a prior is not specified. It is easy in StataStan to add a line of code to the model to include a different prior on $\sigma$ or $\sigma^2$.

A natural hierarchical extension of the Rasch model adds a hyperprior for $\delta_i$ so that,
\begin{equation}
	\mathrm{Pr} (y_{ip} = 1 | \theta_p, \delta_i) =
	\mathrm{logit}^{-1} (\theta_p - \delta_i)
\end{equation}
\begin{equation}
	\theta_p \sim \mathrm{N} (0, \sigma^2)
\end{equation}
\begin{equation}
	\delta_i \sim \mathrm{N} (\mu, \tau^2),
\end{equation}
where $\mu$ is the model intercept. Persons and items are regarded as two sets of exchangeable draws.

\section{Methods}

We simulated data from the above model with $500$ persons each answering  $20$ items. For true values of $\delta_i$, we assigned equally-spaced values from $-1.5$ to $1.5$, and we set the true $\sigma$ to 1.

We set up the Rasch and hierarchical Rasch models in a similar manner, running four chains in series in Stan version 2.11 and Stata version 14.1. We drew initial values for the chains from independent uniform distributions $-1$ to 1 on the location parameters
	$\mu^{(0)}$,
	$\delta^{(0)}$, and
	$\theta^{(0)}$; and uniform distributions from 0 to 2 on the scale parameters
	$\sigma^{(0)}$ and
	$\tau^{(0)}$.
We assigned all $\delta_i$'s identical starting values for each chain, and the same for the $\theta_p$'s. The reason for this (admittedly unusual) choice is that this approach is much easier to employ with \texttt{bayesmh}. We used the same starting values for both StataStan and \texttt{bayesmh} (and in the comparison described below, for JAGS).  These item-response models were not sensitive to starting values.

We ran ten chains for 2500 discarded warm-up iterations and 2500 posterior draws each. For timing purposes, we ran all chains in serial, thus eliminating one of Stan's advantages which is that it can automatically run multiple chains in parallel on a multi-core machine, regardless of the flavor of Stata, although we made one comparison using parallel computation, which is described below. We provide the Stan programs and Stata commands in the appendix. The options specified for {\tt bayesmh} are nearly identical to those in the example provided in the Stata manual \cite{stata14}. There is a difference in how the $\delta$ and $\theta$ parameters are sampled, which plays to the strengths of the different algorithms: Hamiltonian Monte Carlo is more efficient with distributions centered on or close to zero, regardless of correlation, while random walk Metropolis-Hastings in {\tt bayesmh} is improved by using the random effects option \cite{stata14}. This feature, added in Stata 14.1, produces marked improvements in effective sample size for models amenable to a random effects parameterisation. Other model forms will not benefit from it, so for comparison, we ran {\tt bayesmh} both with and without random effects.

We monitored convergence for each parameter using the $\widehat R$ statistic, which is a rough estimate of the square root of the ratio of overall (across chains) posterior variance to within-chain posterior variance \cite{bda3}. Values of $\widehat R$ near $1$ imply convergence, while greater values indicate non-convergence. Values less than $1.1$ are generally considered acceptable. The efficiency of the estimations is evaluated by the seconds per estimated effective sample size, $s / \hat n_{\mathrm{eff}}$  \cite{bda3}. This reflects the fact that more highly autocorrelated chains of draws from the posterior distributions give less precise inference, equivalent to a smaller number of effectively independent samples, which $n_{\mathrm{eff}}$ estimates. Two versions of timings were used: an all-in time using the Stata {\tt timer} from the lines of do-file above and below the {\tt bayesmh} or {\tt stan} command, as well as a simulation-only time, obtained from the CmdStan output and from the value returned to \texttt{e(simtime)} by {\tt bayesmh} (an undocumented return value). StataStan's all-in time includes compilation of the model and {\tt bayesmh}'s includes internal model-building before simulation can begin. To run multiple chains in StataStan, compilation need only happen once, and if data change but a model does not, a previously compiled executable file can be re-used. The total time and simulation-only times represent opposite ends of a spectrum of performance. In real-life implementation, if there are many retained iterations compared to the warm-up iterations, and if compilation (in the case of StataStan) and model building (in the case of {\tt bayesmh}) is not needed in every chain, total time will approach the simulation-only time. 

To further investigate the efficiency of the software as models become more demanding, we carried out the same analyses on simulated data with 20 items and 100, 500, 1000, 5000, and 10,000 people. We compared StataStan 1.2.1 (calling CmdStan 2.11) and Stata 14.1 \texttt{bayesmh} as above, and also the open-source software JAGS 4.0.0 \cite{plummer} via the {\tt rjags} package in R 3.2.3, and ran four chains in each instance. We compared $s / \hat n_{\mathrm{eff}}$ for the hyperparameters $\sigma^2$, $\mu$ and $\tau^2$ and for the worst parameter (lowest $\hat n_{\mathrm{eff}}$, reflecting the frequent need to run the software until all parameters are adequately estimated) in each model. We ran {\tt bayesmh} both with and without the {\tt exclude()} option on the $\theta$s to examine the effect of reducing memory requirements. We also ran StataStan again with parallel chains for 20 items and 1000 people, to examine the increase in speed that is achieved with four CPU cores. All simulations were conducted on an ``Early 2015'' MacBook Pro laptop running OS X 10.11.6 (El Capitan) with a 2.7GHz Intel Core i5 4-core processor and 8GB of 1867MHz DDR3 RAM, with all networking turned off.

\section{Results}

\begin{figure}
	\centerline{
	\includegraphics[width=.49\textwidth]{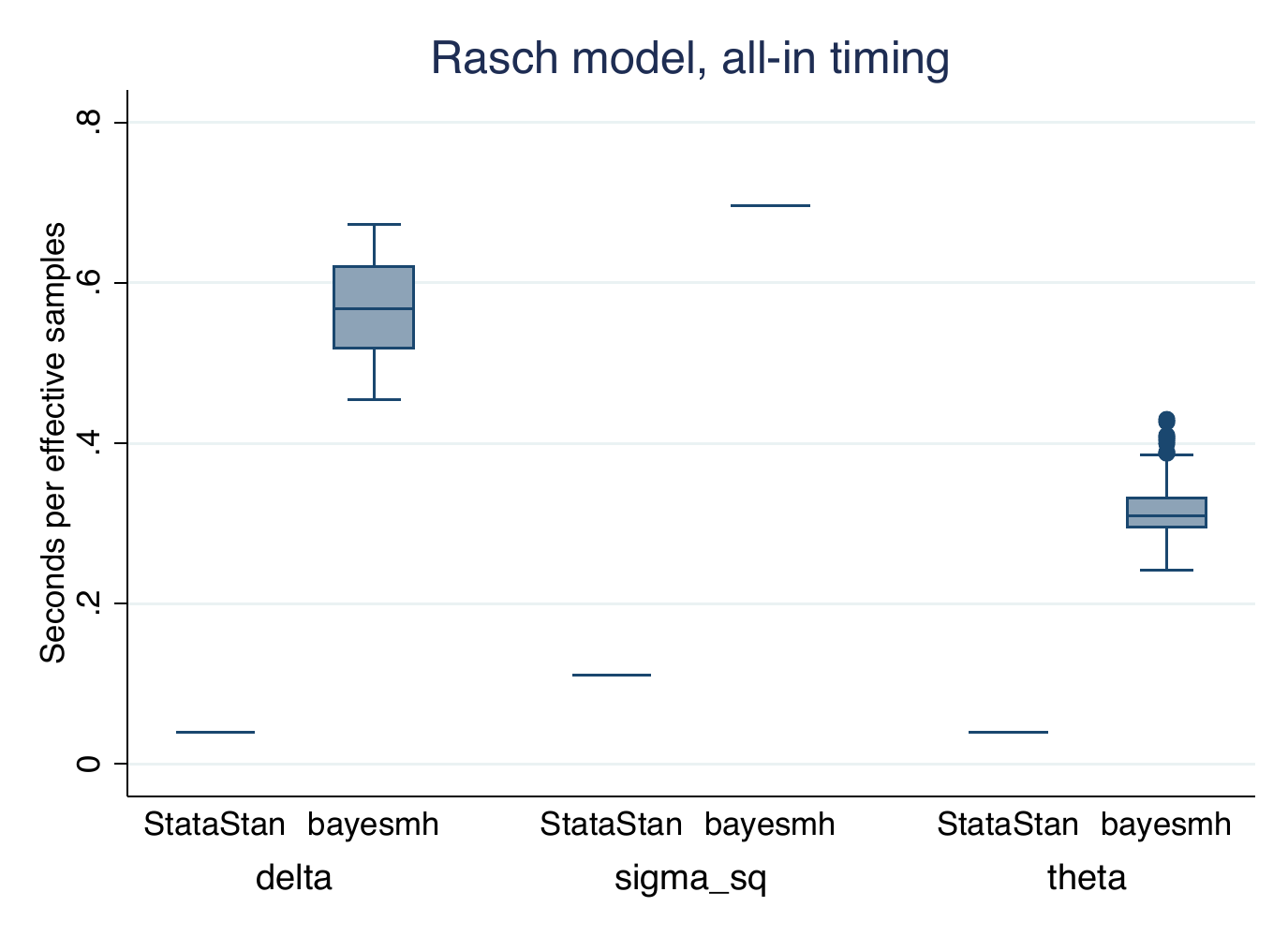}
	\includegraphics[width=.49\textwidth]{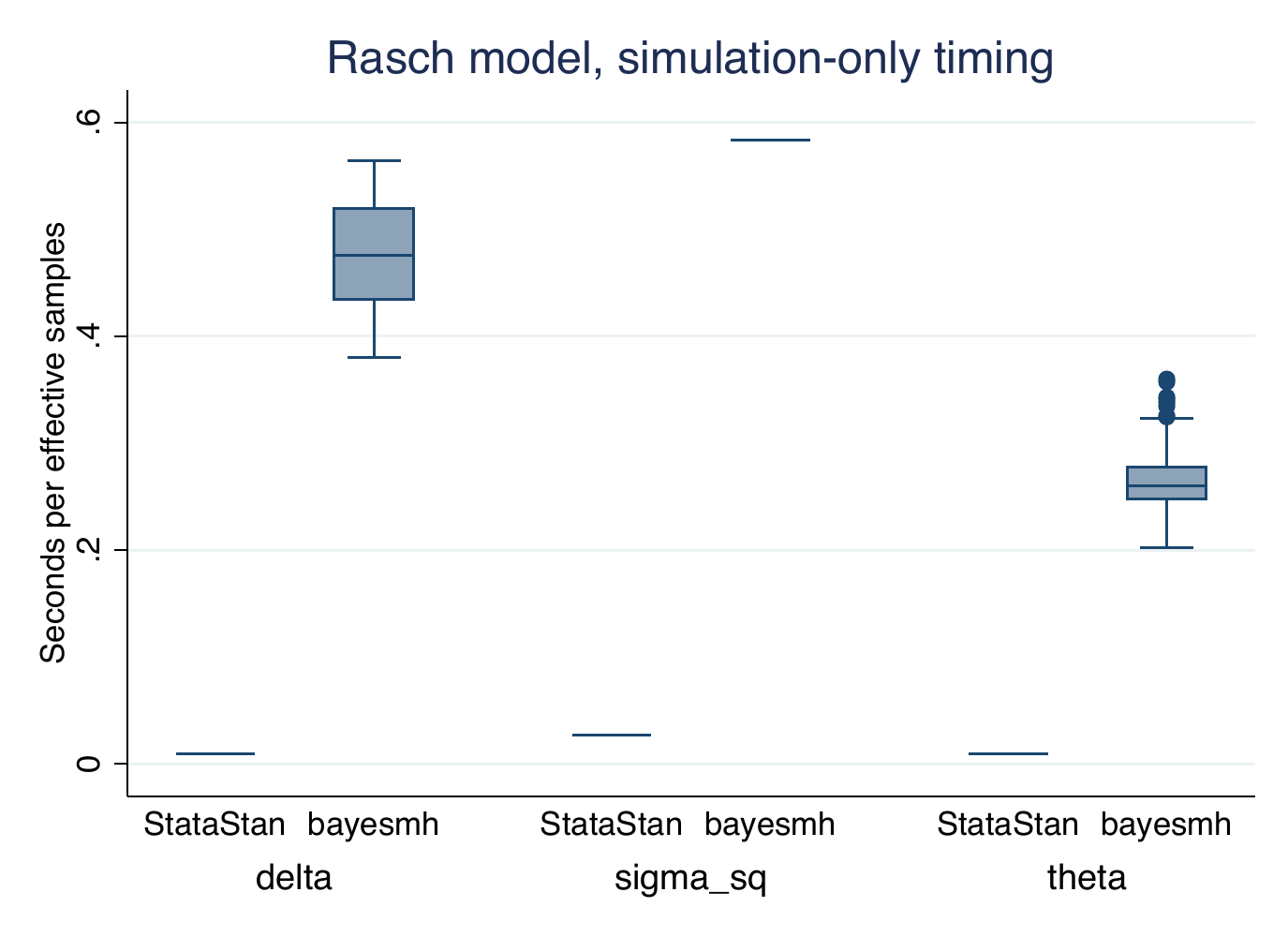}
}
	\centerline{
\includegraphics[width=.49\textwidth]{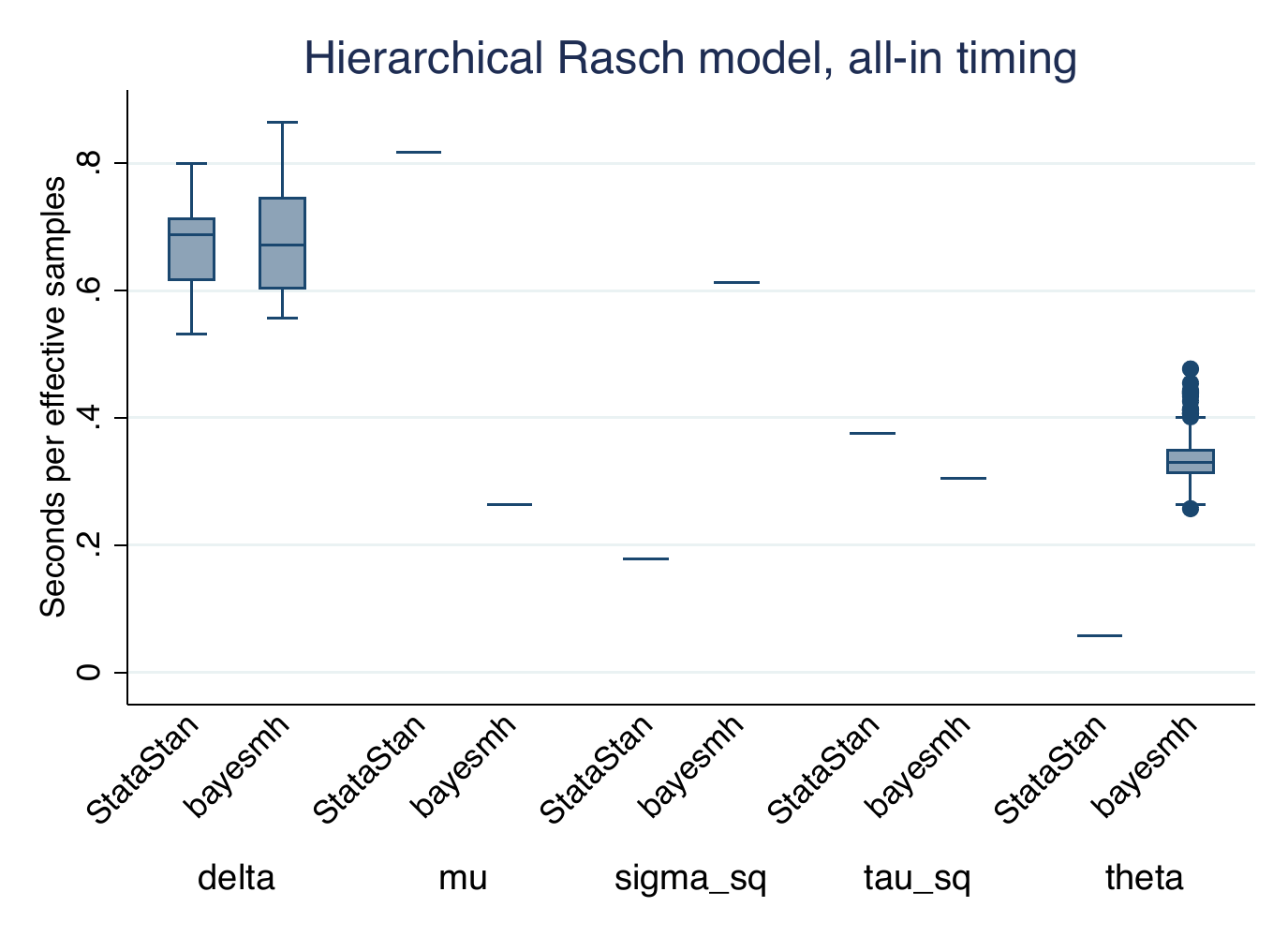}
\includegraphics[width=.49\textwidth]{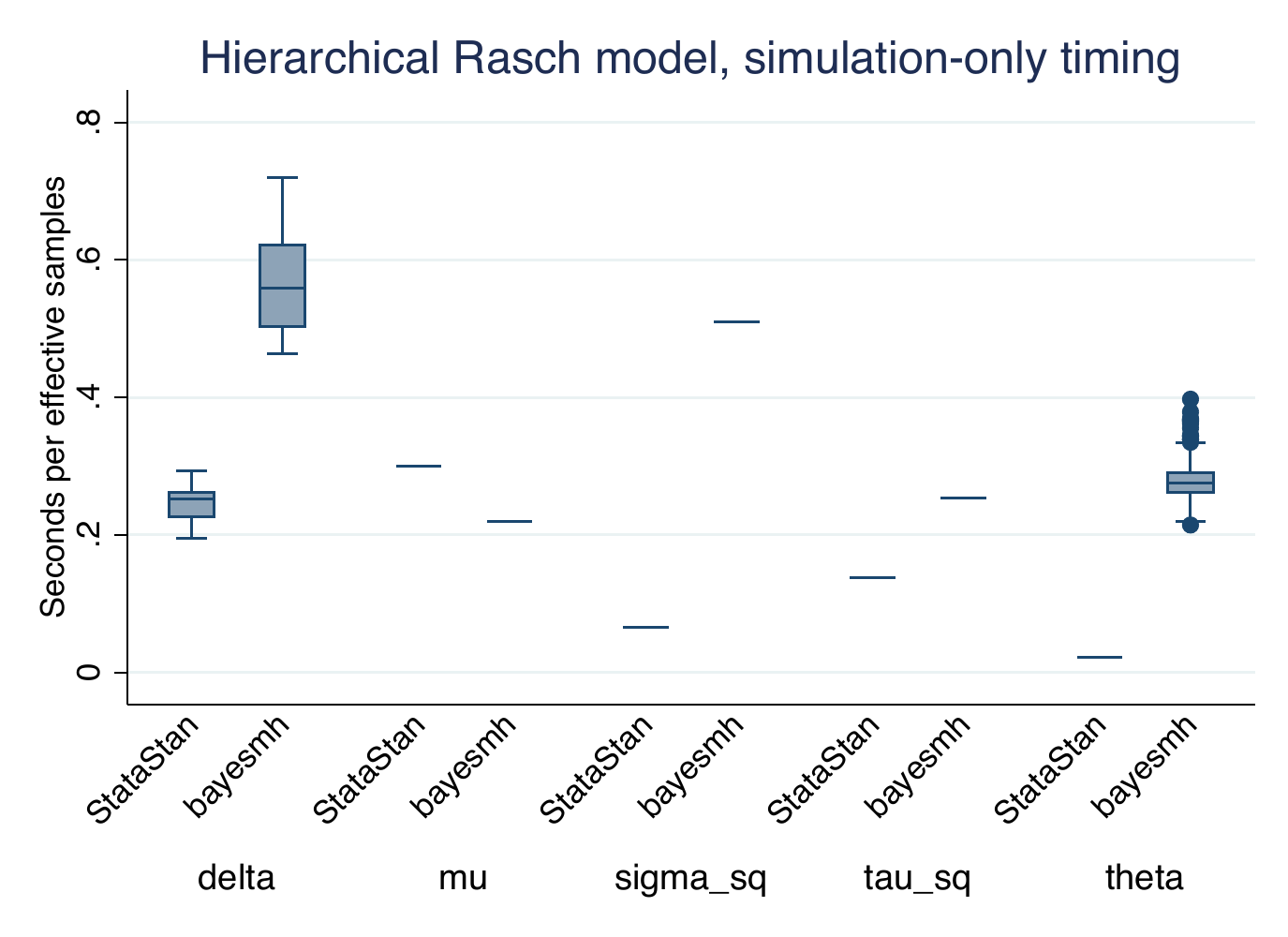}}
	\caption{\em Boxplots of seconds per effective independent sample for parameters in the Rasch model (top row of plots) and hierarchical Rasch model (bottom row), in each case fit to simulated data on 500 persons each answering 20 items. Left column shows total timing including compilation and simulation; right column shows simulation time only. When a model is being fit multiple times, simulation-only timing is a more relevant comparison because the model only needs to be compiled once.}
	\label{fig:box-rasch}
\end{figure}

For the Rasch model, we ran StataStan for ten chains (in series) of 5,000 iterations (first half as warm-up) in \MinutesRaschStan minutes; at that point, $\widehat R$ was less than \UpperRhatRaschStan for all parameters. We ran \texttt{bayesmh} for ten chains of the same length in \MinutesRaschStata minutes; $\widehat R$ was less than \UpperRhatRaschStata for all parameters. Convergence appears satisfactory for both. Values of time per effective independent sample for all the parameters are compared in boxplots (Figure \ref{fig:box-rasch}) between StataStan and {\tt bayesmh}. Table~\ref{tab:table1} provides the same all-in timing statistics for the hyperparameters.

\begin{table}
	\centering
\begin{tabular}{llrr}
  \hline
Model & Parameter & Stata 14.1 bayesmh $n_\mathrm{eff} / \mathrm{sec}$ & StataStan $n_\mathrm{eff} / \mathrm{sec}$ \\ 
  \hline
Rasch & $\sigma^2$ & 1.44 & 8.99  \\ 
  Hierarchical Rasch & $\mu$ & 3.80 & 1.22 \\ 
  Hierarchical Rasch & $\sigma^2$ & 1.63 &  5.62 \\ 
  Hierarchical Rasch & $\tau^2$ & 3.28 & 2.66 \\ 
   \hline
\end{tabular}

	\caption{\em Efficiency statistics for the hyperparameters in the two models.}
	\label{tab:table1}
\end{table}

Results for the hierarchical Rasch model parallel those for the Rasch model. Estimation with StataStan required \MinutesHraschStan minutes for the same number of chains and iterations, and $\widehat R$ was less than \UpperRhatHraschStan for all parameters. \texttt{bayesmh} ran for \MinutesHraschStata minutes and yielded values of $\widehat R$ less than \UpperRhatHraschStata for all parameters. Both estimations appear to have converged. 

In terms of the total time from issuing the command to its completion, StataStan was more efficient for all parameters in the Rasch model, and in the hierarchical Rasch model it was more efficient for all $\theta$s and $\sigma^2$, similar for the $\delta$s, slightly less efficient for $\tau^2$ and less efficient for $\mu$. When we compared simulation-only time (not counting compilation, model building or warm-up), StataStan's efficiency was improved, making all Rasch parameters even more favorable and all hierarchical Rasch parameters except $\mu$ favor StataStan over {\tt bayesmh}.

When we ran the models with the preferred StataStan priors, and sampling standard deviations rather than variances, results were little changed. Total computation time was somewhat faster at 11.1 minutes for the Rasch model and 22.6 minutes for the hierarchical Rasch model, but times per $n_\mathrm{eff}$ were very similar at 0.08 seconds for Rasch $\sigma$, 0.16 seconds for hierarchical Rasch $\sigma$, and 0.29 seconds for $\tau$. However, the efficiency of $\mu$ improved to 0.49 seconds per $n_\mathrm{eff}$.

\begin{table}
	\centering
\begin{tabular}{lcccccc}
& & & \multicolumn{2}{|c|}{Total time} & \multicolumn{2}{|c|}{Simulation-only time} \\
& & & bayesmh & StataStan & bayesmh & StataStan \\
Model & Parameter & $P$ & $\mathrm{sec} / n_\mathrm{eff}$ & $\mathrm{sec} / n_\mathrm{eff}$ & $\mathrm{sec} / n_\mathrm{eff}$ & $\mathrm{sec} / n_\mathrm{eff}$ \\
\hline
Rasch & $\sigma^2$ & 100 &  0.143 &  0.069 &  0.137 &  0.007 \\
 & & 500 &  0.536 &  0.105 &  0.502 &  0.025 \\
 & & 1000 &  1.460 &  0.230 &  1.319 &  0.062 \\
 & & 5000 &  9.333 &  1.649 &  6.404 &  0.576 \\
 & & 10000 & 350.164 &  4.539 & 334.916 &  1.487 \\
\hline
H. Rasch & $\mu$ & 100 &  0.212 &  0.168 &  0.204 &  0.023 \\
 & & 500 &  0.211 &  0.760 &  0.197 &  0.287 \\
 & & 1000 &  0.457 &  1.131 &  0.413 &  0.571 \\
 & & 5000 &  2.682 & 22.025 &  1.847 & 11.331 \\
 & & 10000 & 49.533 & 67.812 & 46.660 & 37.400 \\
\hline
H. Rasch & $\sigma^2$ & 100 &  0.146 &  0.061 &  0.140 &  0.008 \\
 & & 500 &  0.595 &  0.177 &  0.558 &  0.067 \\
 & & 1000 &  1.809 &  0.340 &  1.634 &  0.172 \\
 & & 5000 & 11.941 &  4.508 &  8.225 &  2.319 \\
 & & 10000 & 186.637 & 13.236 & 175.813 &  7.300 \\
\hline
H. Rasch & $\tau^2$ & 100 &  0.094 &  0.095 &  0.090 &  0.013 \\
 & & 500 &  0.350 &  0.385 &  0.328 &  0.145 \\
 & & 1000 &  0.904 &  0.608 &  0.817 &  0.307 \\
 & & 5000 &  5.145 &  8.237 &  3.544 &  4.237 \\
 & & 10000 & 76.556 & 26.884 & 72.116 & 14.827 \\
\hline
\end{tabular}

	\caption{\em Efficiency statistics for hyperparameters in increasingly large models.}
	\label{tab:table2}
\end{table}

\begin{figure}
	\centering	\includegraphics[width=.51\textwidth]{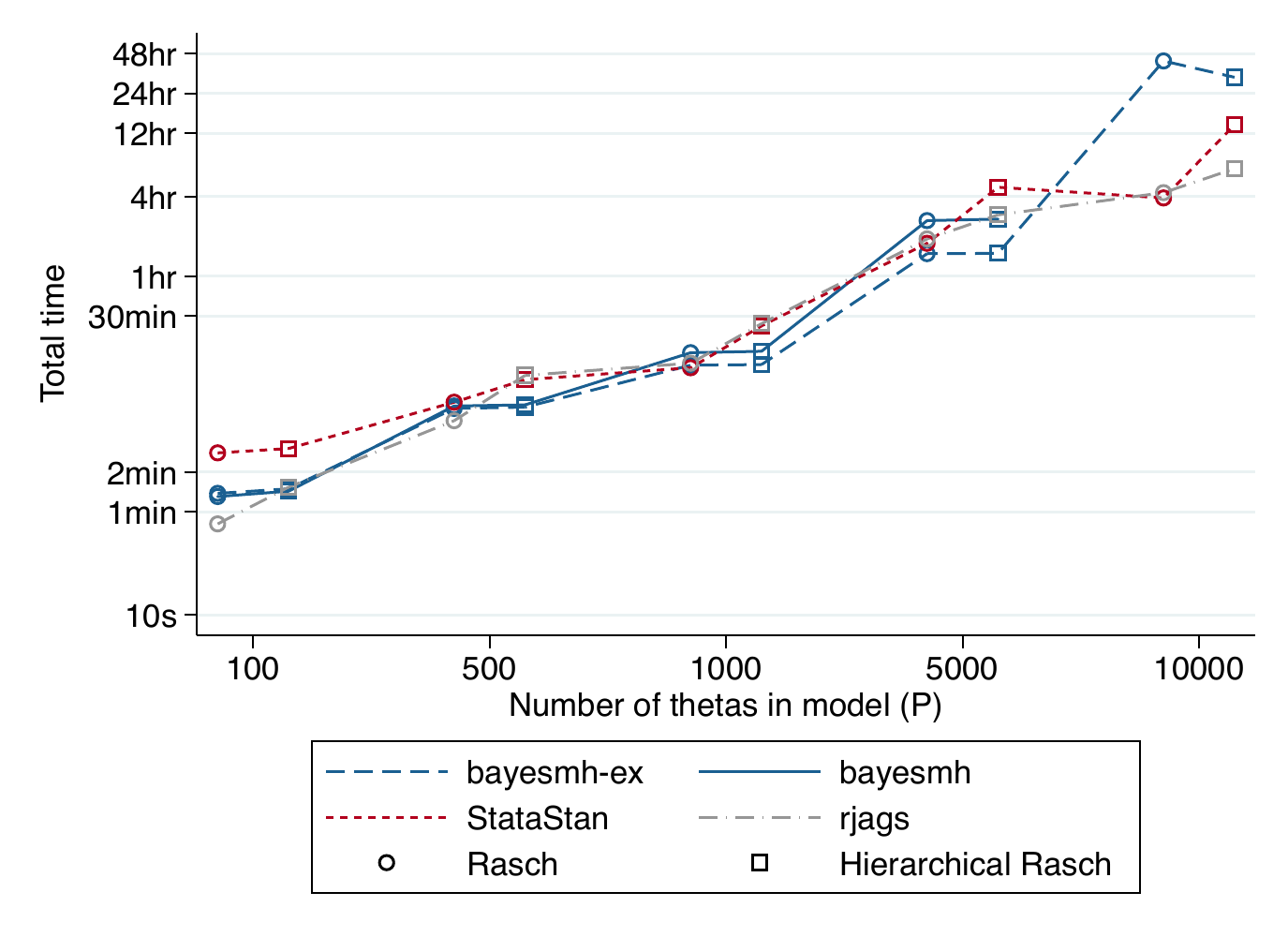}
	\caption{\em Total time per effective independent sample (worst efficiency across all parameters) in increasingly large Rasch and hierarchical Rasch models.}
	\label{fig:totaltime}
\end{figure}
\begin{figure}
	\centering	\includegraphics[width=.51\textwidth]{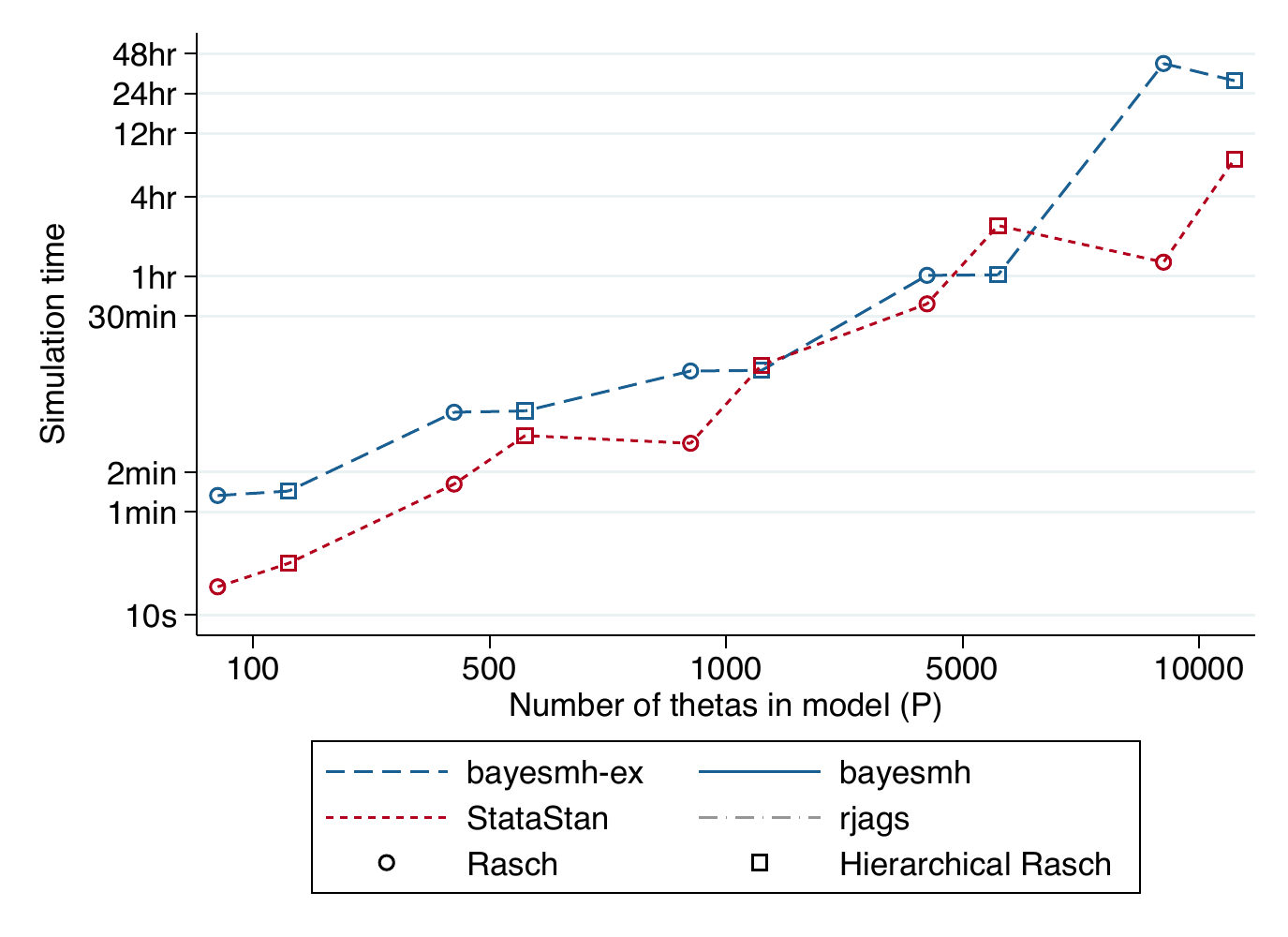}
	\caption{\em Total time per effective independent sample (worst efficiency across all parameters) in increasingly large Rasch and hierarchical Rasch models.}
	\label{fig:simtime}
\end{figure}

In testing with increasingly large models, all three packages showed similar total execution times and StataStan had faster simulation-only time in all the Rasch models (from 3\% to 61\% of the {\tt bayesmh} times) and mixed results in the hierarchical Rasch models (from 26\% to 235\%). The charts show the efficiency for the parameter with the lowest $n_\mathrm{eff}$ in each case, for total time (Figure \ref{fig:totaltime}) and simulation-only time (Figure \ref{fig:simtime}). In these line charts, the form of {\tt bayesmh} that uses the {\tt exclude} option is denoted by ``bayesmh-ex''; we found this had similar speed and efficiency to {\tt bayesmh} without {\tt exclude} and so did not assess it further in the most time-consuming models ($P=10,000$) or in terms of simulation-only time. JAGS does not provide simulation-only timings. In total time per $n_\mathrm{eff}$, no one software option dominated, though from this limited simulation, it appeared that StataStan was more efficient in the smallest models and \texttt{bayesmh} in the largest (Figure \ref{fig:totaltime}). In simulation-only time per $n_\mathrm{eff}$, StataStan was more efficient than \texttt{bayesmh}, up to 10 times so, in most models and sizes of $P$, but in some cases they were similar with \texttt{bayesmh} being slightly better (Figure \ref{fig:simtime}).

StataStan consistently had the better efficiency ($s / n_\mathrm{eff}$) for $\sigma^2$ in both Rasch and hierarchical Rasch models, but mixed results for $\tau^2$, although 4 out of 5 models favoured StataStan in simulation-only time (Table \ref{tab:table2}). In the Rasch model, StataStan was 2.1 to 77.1 times faster than \texttt{bayesmh} to achieve the same $n_\mathrm{eff}$ for $\sigma^2$ in total time, and 11.1 to 225.2 times faster in simulation-only time. In the hierarchical Rasch model, StataStan was 2.4 to 14.1 times faster for $\sigma^2$ in total time and 3.5 to 24.1 times in simulation-only time. StataStan was 0.6 to 2.8 times faster for $\tau^2$ in total time and 0.8 to 6.9 times faster in simulation-only time. The $\mu$ hyperparameter in the hierarchical Rasch models was more efficiently sampled by \texttt{bayesmh} at most values of $P$, with StataStan being 0.1 to 1.3 times faster in total time and 0.2 to 8.9 times faster in simulation-only time (Table \ref{tab:table2}). All models, with all software, could be fitted with the same laptop computer without running out of memory.

The random effects option in {\tt bayesmh} provided a considerable improvement in both effective sample size and speed. When we ran the $I=20, P=100$ models without random effects, total time was 206 seconds for Rasch and 211 seconds for hierarchical Rasch, while simulation-only times were 200 and 204 seconds respectively, which is about 2.5 times slower than the same model with random effects. The time per effective independent sample was considerably increased. In the Rasch model, it rose from 0.143 to 69 seconds of $\sigma^2$. In the hierarchical Rasch model, it rose from 0.146 to 30 seconds for $\sigma^2$, from 0.212 to 53 seconds for $\mu$, and from 0.094 to 23 seconds for $\tau^2$.

A further consideration is the speed-up obtained by running StataStan chains in parallel even without Stata/MP. We found that the Rasch model with $I=20$ and $P=1000$ had total time 383 seconds running in parallel compared to 734 seconds running in series, and simulation-only time of 78 seconds compared to 198 seconds. The hierachical Rasch model of the same size had total time 850 seconds compared to 1520 seconds and simulation-only time of 303 seconds compared to 768 seconds. This would make parallel StataStan on a quad-core computer roughly twice as efficient as serial StataStan, while {\tt bayesmh} will not run parallel chains without Stata/MP.

\section{Discussion}

We found that most of the Rasch models we compared were more efficiently sampled by StataStan than \texttt{bayesmh}, and that this was more favourable to StataStan as the fixed overhead of compiling the model into an executable file was outgrown by the simulation. This suggests that longer chains of draws from the posterior distribution, pre-compiled models, and parallel chains will all favour StataStan, and the total time comparisons here represent a worst case scenario for StataStan. We would expect the results we found to apply to generalized linear mixed models, given that these include the Rasch models as a special case \cite{rijmen2003nonlinear, zheng2007estimating}. In practice, the adaptive Markov chain Monte Carlo algorithm featured in {\tt bayesmh} (Stata v14.1) also has a number of features that improve its performance notably over JAGS or Stata v14.0. We found that the same models without the {\tt reffects} option took 200 to 500 times longer to achieve the same effective sample size on {\tt bayesmh}, which should be borne in mind when considering models outside the Rasch family, when random effects are not used.

StataStan provides a simple interface, operating by writing specified variables (as vectors), matrices and scalars from Stata to a text file and calling the command-line implementation of Stan. A wide variety of priors can be specified, and the algorithm is less sensitive to the prior than that used in \texttt{bayesmh}. In these Rasch models, we found it simple and more intuitive to sample the hyperparameters as standard deviations rather than variances or precisions, and we employed uniform priors as the Stan default without any impact on efficiency. The alternative programs are given in the appendix. Progress is displayed inside Stata (even under Windows) and there is the option to write the Stan model inside a comment block in the Stata do-file. Results can then be read back into Stata for diagnostics, generating other values of interest, or saving in \texttt{.dta} format. StataStan can be installed from SSC by 

\texttt{ssc install stan}

\noindent Windows users should also run:

\texttt{ssc install windowsmonitor}

\noindent In conclusion, we find StataStan to be generally faster than \texttt{bayesmh}, which is no surprise given Stan's advanced algorithms and efficient autodifferentiation code. Given that Stan is open-source, offers a wider range of models than \texttt{bayesmh}, and can be run directly from Stata using StataStan, we recommend that Stata users who are interested in Bayesian methods investigate StataStan and consider using it for Bayesian modelling, especially for more complicated models.


\begin{thebibliography}{99}

\bibitem{stata14}
StataCorp. \emph{Stata Statistical Software: Release 14.1}. (2016) College Station, TX.

\bibitem{StataStan}
Robert L Grant. \emph{StataStan}. (2015) https://github.com/stan-dev/statastan Accessed 4 October 2016.

\bibitem{stanmanual}
Stan Development Team. \emph{Stan Modeling Language: User's Guide and Reference Manual}. (2016) http://mc-stan.org/documentation/ Accessed 4 October 2016.

\bibitem{hoffman}
Matthew Hoffman and Andrew Gelman. The no-U-turn sampler: Adaptively setting path lengths in Hamiltonian Monte Carlo. \emph{Journal of Machine Learning Research} (2014);15:1593--1623.

\bibitem{kucukelbir}
Alp Kucukelbir, Rajesh Ranganath, Andrew Gelman and David Blei. Automatic Variational Inference in Stan. \emph{ArXiv} (2015):1506.03431.

\bibitem{nocedal}
Jorge Nocedal and Stephen Wright. \emph{Numerical Optimization}, 2nd edition (2006). Springer-Verlag, Berlin.

\bibitem{neal}
Radford Neal. MCMC using Hamiltonian dynamics. In \emph{Handbook of Markov Chain Monte Carlo}, eds S Brooks, A Gelman, GL Jones and X-L Meng. (2011) CRC Press, Boca Raton, FL.

\bibitem{rasch1960}
Georg Rasch (1960). \emph{Probabilistic Models for Some Intelligence and Achievement Tests}. University of Chicago Press, Chicago.

\bibitem{priors}
Andrew Gelman. Prior distributions for variance parameters in hierarchical models. \emph{Bayesian Analysis} (2006);1:515--533.

\bibitem{bda3}
Andrew Gelman, John B. Carlin, Hal S. Stern, David B. Dunson, Aki Vehtari and Donald B. Rubin (2013). \emph{Bayesian Data Analysis}, 3rd edition. CRC Press, Boca Raton, FL.

\bibitem{plummer}
Martyn Plummer. \emph{JAGS: Just Another Gibbs Sampler}. (2007) http://mcmc-jags.sourceforge.net/ Accessed 4 October 2016.

\bibitem{zheng2007estimating}
Xiaohui Zheng and Sophia Rabe-Hesketh. Estimating parameters of dichotomous and ordinal item response models with \texttt{gllamm}. \emph{Stata Journal} (2007);7(3):313--333.

\bibitem{rijmen2003nonlinear}
Frank Rijmen, Francis Tuerlinckx, Paul De Boeck and Peter Kuppens. A nonlinear mixed model framework for item response theory. \emph{Psychological Methods} (2003);8(2):185--205.



\end{thebibliography}

\section*{Appendix}

Here is the code for the models, starting with the Rasch Stan program which matches \texttt{bayesmh}:

\begin{small}
\begin{verbatim}
data {
  int<lower=1> N; // number of observations in the dataset
  int<lower=1> I; // number of items
  int<lower=1> P; // number of people
  int<lower=1, upper=I> ii[N]; // variable indexing the items
  int<lower=1, upper=P> pp[N]; // variable indexing the people
  int<lower=0, upper=1> y[N]; // binary outcome variable
}
parameters {
  real<lower=0> sigma_sq; // variance of the thetas (random intercepts for people)
  vector[I] delta_unit; // normalised deltas
  vector[P] theta_unit; // normalised thetas
}
transformed parameters {
  real<lower=0> sigma;
  sigma = sqrt(sigma_sq); // SD of the theta random intercepts
}
model {
  vector[I] delta;
  vector[P] theta;
  theta_unit ~ normal(0, 1); // prior for normalised thetas
  delta_unit ~ normal(0, 1); // prior for normalised deltas
  sigma_sq ~ inv_gamma(1, 1); // prior for variance of thetas 
  theta = theta_unit * sigma; // convert normalised thetas to thetas (mean 0)
  delta = delta_unit * sqrt(10); // convert normalised deltas to deltas (mean 0)
  y ~ bernoulli_logit(theta[pp] - delta[ii]); // likelihood
}
\end{verbatim}
\end{small}

This is our preferred Stan program:

\begin{small}
\begin{verbatim}
data {
  int<lower=1> N; // number of observations in the dataset
  int<lower=1> I; // number of items
  int<lower=1> P; // number of people
  int<lower=1, upper=I> ii[N]; // variable indexing the items
  int<lower=1, upper=P> pp[N]; // variable indexing the people
  int<lower=0, upper=1> y[N]; // binary outcome variable
}
parameters {
  real<lower=0> sigma; // SD of the thetas (random intercepts for people)
  vector[I] delta_unit; // normalised deltas
  vector[P] theta_unit; // normalised thetas
}
model {
  vector[I] delta;
  vector[P] theta;
  theta_unit ~ normal(0, 1); // prior for normalised thetas
  delta_unit ~ normal(0, 1); // prior for normalised deltas
  theta = theta_unit * sigma; // convert normalised thetas to thetas (mean 0)
  delta = delta_unit * sqrt(10); // convert normalised deltas to deltas (mean 0)
  y ~ bernoulli_logit(theta[pp] - delta[ii]); // likelihood
}
\end{verbatim}
\end{small}

Here is the Stata call for the Rasch model:

\begin{small}
\begin{verbatim}
bayesmh y=({theta:}-{delta:}), likelihood(logit) ///
   redefine(delta:i.item) redefine(theta:i.person) ///
   prior({theta:i.person}, normal(0, {sigmasq})) ///
   prior({delta:i.item}, normal(0, 10)) ///
   prior({sigmasq}, igamma(1, 1)) ///
   mcmcsize(`mcmcsize') burnin(`burnin') ///
   notable saving(`draws', replace) dots ///
   initial({delta:i.item} `=el(inits`jj', `c', 1)'  ///
      {theta:i.person} `=el(inits`jj', `c', 2)' ///
      {sigmasq} `=el(inits`jj', `c', 3)') ///
   block({sigmasq}) 
\end{verbatim}
\end{small}

And here is the JAGS code for the Rasch model:

\begin{small}
\begin{verbatim}
model {
  for (i in 1:I) {
    delta[i] ~ dunif(-1e6, 1e6)
  }
  inv_sigma_sq ~ dgamma(1,1)
  sigma <- pow(inv_sigma_sq, -0.5)
  for (p in 1:P) {
    theta[p] ~ dnorm(0, inv_sigma_sq)
  }
  for (n in 1:N) {
    logit(inv_logit_eta[n]) <- theta[pp[n]] - delta[ii[n]]
    y[n] ~ dbern(inv_logit_eta[n])
  }
}
\end{verbatim}
\end{small}

Here is the hierarchical Rasch model in Stan, matching \texttt{bayesmh}:

\begin{small}
\begin{verbatim}
data {
  int<lower=1> N; // number of observations in the dataset
  int<lower=1> I; // number of items
  int<lower=1> P; // number of people
  int<lower=1, upper=I> ii[N]; // variable indexing the items
  int<lower=1, upper=P> pp[N]; // variable indexing the people
  int<lower=0, upper=1> y[N]; // binary outcome variable
}
parameters {
  real<lower=0> sigma_sq; // variance of the thetas (random intercepts for people)
  real<lower=0> tau_sq; // variance of the deltas (random intercepts for items)
  real mu; // mean of the deltas
  vector[I] delta_unit; // normalised deltas
  vector[P] theta_unit; // normalised thetas
}
transformed parameters {
  real<lower=0> sigma;
  real<lower=0> tau;
  sigma = sqrt(sigma_sq); // SD of the theta random intercepts
  tau = sqrt(tau_sq); // SD of the delta random intercepts
}
model {
  vector[I] delta;
  vector[P] theta;
  theta_unit ~ normal(0, 1); // prior for normalised thetas
  delta_unit ~ normal(0, 1); // prior for normalised deltas
  mu ~ normal(0, sqrt(10)); // prior for the mean of the deltas
  sigma_sq ~ inv_gamma(1, 1);
  tau_sq ~ inv_gamma(1, 1);
  theta = theta_unit * sigma; // convert normalised thetas to thetas (mean 0)
  delta = mu + (delta_unit * tau); // convert normalised deltas to deltas (mean mu)
  y ~ bernoulli_logit(theta[pp] - delta[ii]); // likelihood
}
\end{verbatim}
\end{small}

This is our preferred Stan model:

\begin{small}
\begin{verbatim}
data {
  int<lower=1> N; // number of observations in the dataset
  int<lower=1> I; // number of items
  int<lower=1> P; // number of people
  int<lower=1, upper=I> ii[N]; // variable indexing the items
  int<lower=1, upper=P> pp[N]; // variable indexing the people
  int<lower=0, upper=1> y[N]; // binary outcome variable
}
parameters {
  real<lower=0> sigma; // SD of the thetas (random intercepts for people)
  real<lower=0> tau; // SD of the deltas (random intercepts for items)
  real mu; // mean of the deltas
  vector[I] delta_unit; // normalised deltas
  vector[P] theta_unit; // normalised thetas
}
model {
  vector[I] delta;
  vector[P] theta;
  theta_unit ~ normal(0, 1); // prior for normalised thetas
  delta_unit ~ normal(0, 1); // prior for normalised deltas
  mu ~ normal(0, sqrt(10)); // prior for the mean of the deltas
  theta = theta_unit * sigma; // convert normalised thetas to thetas (mean 0)
  delta = mu + (delta_unit * tau); // convert normalised deltas to deltas (mean mu)
  y ~ bernoulli_logit(theta[pp] - delta[ii]); // likelihood
}
\end{verbatim}
\end{small}

Here is the Stata call for the hierarchical Rasch model:

\begin{small}
\begin{verbatim}
bayesmh y=({theta:}-{delta:}),likelihood(logit) ///
   redefine(delta:i.item) redefine(theta:i.person) ///
   prior({theta:i.person}, normal(0, {sigmasq})) ///
   prior({delta:i.item}, normal({mu}, {tausq})) ///
   prior({mu}, normal(0, 10)) ///
   prior({sigmasq} {tausq}, igamma(1, 1)) ///
   block({sigmasq} {tausq} {mu}, split) ///
   initial({delta:i.item} `=el(inits`jj', 1, 1)' ///
      {theta:i.person} `=el(inits`jj', 1, 2)' ///
      {sigmasq} `=el(inits`jj', 1, 3)' ///
      {tausq} `=el(inits`jj', 1, 4)' ///
      {mu} `=el(inits`jj', 1, 5)') ///
   mcmcsize(`mcmcsize') burnin(`burnin') ///
   saving(`draws', replace) dots
\end{verbatim}
\end{small}
And here is the JAGS code for the hierarchical Rasch model:
\begin{small}
\begin{verbatim}
model {
  inv_sigma_sq ~ dgamma(1,1)
  sigma <- pow(inv_sigma_sq, -0.5)
  for (p in 1:P) {
    theta[p] ~ dnorm(0, inv_sigma_sq)
  }
  inv_tau_sq ~ dgamma(1,1)
  tau <- pow(inv_tau_sq, -0.5)
  for (i in 1:I) {
    delta[i] ~ dnorm(0, inv_tau_sq)
  }
  mu ~ dunif(-1e6, 1e6)
  for (n in 1:N) {
    logit(inv_logit_eta[n]) <- mu + theta[pp[n]] - delta[ii[n]]
    y[n] ~ dbern(inv_logit_eta[n])
  }
}
\end{verbatim}
\end{small}

\end{document}